\documentclass{sig-alternate}

\usepackage{verbatim} 

\begin{document}

\conferenceinfo{CIDR}{ 2007 Asilomar, CA USA}

\title{bdbms -- A Database Management System for Biological Data}

\numberofauthors{1}

\author{
\alignauthor Mohamed Y. Eltabakh~~~~Mourad Ouzzani~~~~Walid G. Aref \\
       \affaddr{Department of Computer Science }\\
       \affaddr{Purdue University }\\
       \affaddr{West Lafayette, IN}\\
       \email{\{meltabak,mourad,aref\}@cs.purdue.edu}
}

\maketitle

\begin{abstract}
Biologists are increasingly using databases for storing and managing their data. Biological
databases typically consist of a mixture of raw data, metadata, sequences, annotations, and
related data obtained from various sources. Current database technology
lacks several functionalities that are needed by biological databases.
In this paper, we introduce {\em bdbms}, an extensible prototype database management system for supporting biological data.
bdbms extends the functionalities of current DBMSs with: (1) Annotation and provenance management
including storage, indexing, manipulation, and querying of annotation and provenance as first class objects in bdbms,
(2) Local dependency tracking to track the dependencies and derivations among data items,
(3) Update authorization to support data curation via {\em content-based} authorization,
in contrast to identity-based authorization, and
(4) New access methods and their supporting operators
that support pattern matching on various types of compressed biological data types.
This paper presents the design of bdbms along with
the techniques proposed to support these functionalities including an extension to SQL.
We also outline some open issues in building bdbms.
\end{abstract}

\section{Introduction}
\begin{sloppypar}
Biological databases are essential to biological experimentation and analysis. They are 
used at different stages of life science research to deposit raw data, store 
interpretations of experiments and results of analysis processes, and search for matching 
structures and sequences. As such, they represent the backbone of life sciences 
discoveries. However, current database technology has not kept pace with the proliferation and 
specific requirements of biological databases~\cite{p21, p20}. In fact, the limited ability of database 
engines to furnish the needed functionalities to manage and process biological data 
properly has become a serious impediment to scientific progress. 

In many cases, biologists tend to store their data in flat files  or spreadsheets  mainly because 
current database systems lack several functionalities that are needed by 
biological databases, e.g., efficient support for sequences, annotations, and provenance.  
Once the data resides outside a 
database system, it loses effective and efficient manageability. Consequently, many of the advantages and functionalities that  
database systems offer are nullified and bypassed. It is thus important to break this 
inefficient and ineffective cycle by empowering database engines to operate directly on the data 
from within its natural habitat; the database system.

Biological databases evolve in an environment with rapidly changing experimental technologies
and  semantics of the information content
and also in a social context that lacks absolute authority to verify correctness of information.  
Furthermore, because the only authority is the scientific community itself, biological databases often require some 
form of community-based curation. These characteristics make it difficult, even using good design strategies, 
to completely foresee the kinds of additional information (termed annotations) that, over time, may become necessary 
to attach to data in the database. 

In this paper, we propose bdbms, an extensible prototype database engine 
for supporting and processing biological databases. While there are several functionalities of interest, 
we focus on the following key features: (1) Annotation and provenance management, 
(2) Local dependency tracking, (3) Update authorization, and 
(4) Non-traditional and novel access methods.   
bdbms will make fundamental advances in the use of biological databases 
through new native and transparent support mechanisms at the DBMS level.

Annotations and provenance data are treated as first-class objects inside bdbms. bdbms provides a framework 
that allows adding annotations/provenance at multiple granularities, i.e., table, tuple, column, and cell levels, archiving and restoring 
annotations, and querying the data based on the annotation/provenance values. In bdbms, we introduce an extension to SQL, termed Annotation-SQL, or {\em A-SQL} for short, to support the processing and querying 
of annotation and provenance information. A-SQL allows annotations and provenance data to be seamlessly propagated 
with query answers with minimal user programming. 

\begin{figure}
 \centering
   \includegraphics[height=2.5cm, width= 8cm, angle=0]{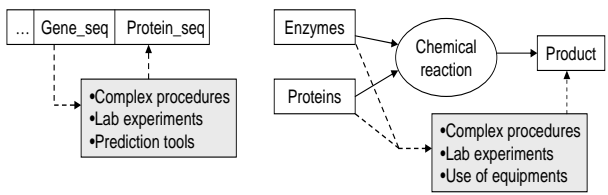}
      \caption{Local dependencies}
    \label{fig:dependency-example}
\end{figure}

In bdbms, we propose a systematic approach for tracking dependencies among database items. 
As a result, when a database item is modified, bdbms can track and mark any other item that is affected by this modification and that needs to be re-verified. 
This feature is particularly desirable in biological databases because many dependencies cannot be computed using coded functions. 
For example (refer to Figure~\ref{fig:dependency-example}), 
protein sequences are derived from gene (DNA) sequences. If a gene sequence is modified, the corresponding protein sequence(s), 
derived calculated quantities (such as molecular weight), and annotations may become invalid. 
Similarly, we may store descriptions of chemical reactions, e.g., substrates, reaction parameters, and products. 
If any of the substrates in the reaction are modified, the products of the reaction are likely to require re-evaluation. 
However, since these dependencies are complex and involve lab experiments and external analysis, 
database systems cannot systematically re-compute the other affected items. 
Lack of system support to automatically track such dependencies raises significant concerns on the quality and the consistency of the data maintained in biological databases.

Authorizing database operations is also one of the features that we extended in bdbms. 
Current database systems support the GRANT/REVOKE access models that depend only on the identity of the user. 
In bdbms, we propose the concept of {\em content-based} authorization, i.e., 
the authorization is based not only on the identity of an updater but also on the content of the updated data. 
For example, lab members may have the authority to update a given data set. However, for credibility and reliability of the data, these updates have to be revised by the lab 
administrator. The lab administrator can then approve or disapprove the operations based on their contents. 
In the mean time, users may be allowed to view the data pending its approval/disapproval.

The other key feature in bdbms is to provide access methods for supporting various types of biological data. 
Our goal is to design and integrate non-traditional and novel access methods inside bdbms.  
For example, sequences and multi-dimensional data are very common in biological databases and hence there is a need to
integrate new types of index structures such as SP-GiST~\cite{p2, p1, p4, p3}
and the SBC-tree~\cite{p5} inside bdbms along with their supporting operators.
SP-GiST is an extensible indexing framework for supporting multi-dimensional data while the SBC-tree is an index structure for indexing and querying compressed sequences 
without decompressing them.  

The rest of the paper is organized as follows. In Section~\ref{sec:arch}, we present the overall architecture of bdbms. 
In Sections~\ref{sec:annotation}--\ref{sec:indexing}, we present each of the bdbms key features. 
Section~\ref{sec:relatedwork} overviews the related work, and Section~\ref{sec:conclusion} contains concluding remarks.
\end{sloppypar}

\section{bdbms System Architecture}
\label{sec:arch}

\begin{sloppypar}
The main components of bdbms are the annotation manager, the dependency manager, 
and the authorization manager.
{\em A-SQL} is bdbms's extended SQL that  
supports annotation (Section~\ref{sec:annotation})  
and authorization commands (Section~\ref{sec:approval}).  
bdbms's {\em annotation manager} is responsible for handling the annotations in an {\em annotation storage} space 
(Section~\ref{sec:annotation}). 
The {\em dependency manager} is responsible for handling the dependencies and derivations among database items. These dependencies are stored in 
a {\em dependency storage} space (Section~\ref{sec:dependency}).  
The {\em authorization manager} handles {\em content-based} authorizations as well as the standard GRANT/REVOKE authorizations over 
the database (Section~\ref{sec:approval}). 
{\em Index structures} are available in bdbms in support of the multidimensional and compressed data (Section~\ref{sec:indexing}).     
\end{sloppypar}

\section{Annotation Management}
\label{sec:annotation}

\begin{figure*}
 \centering
   \includegraphics[height=17cm, width= 4cm, angle=270]{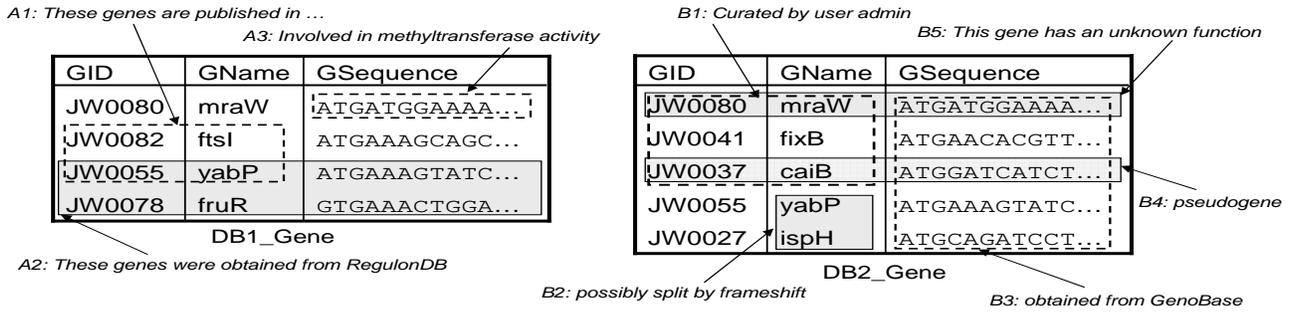}
      \caption{Annotating tables DB1\_Gene and DB2\_Gene}
    \label{fig:ann-example1}
\end{figure*}

\begin{figure*}
 \centering
   \includegraphics[height=17cm, width= 3cm, angle=270]{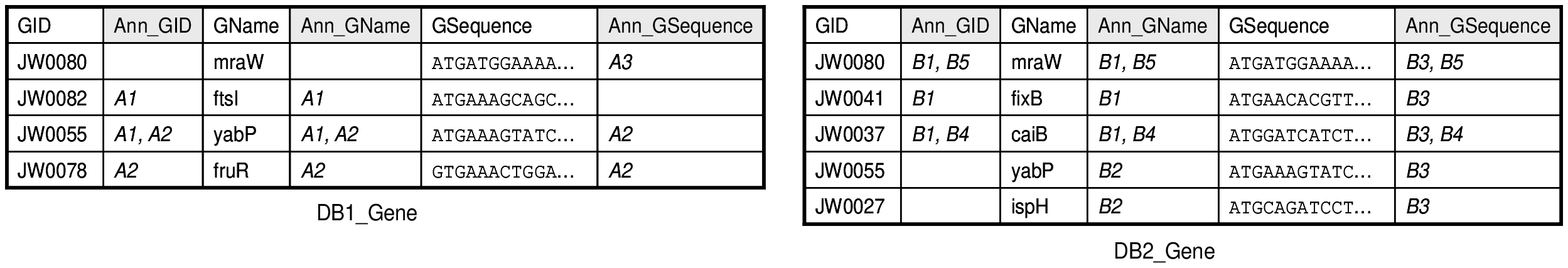}
      \caption{Simple annotation storage scheme: Every data column has a corresponding annotation column}
    \label{fig:ann-example2}
\end{figure*}

\begin{sloppypar}
Annotations are extra information linked to data items inside a database. 
They usually represent users' comments, experiences, related information that is not modeled by the database schema,  
or the provenance (lineage) of the data.  
Adding and retrieving annotations represent an important way of communication and interaction among database users. 
In biological databases, annotations are used extensively to allow users to have a better understanding of the data, e.g.,
how a piece of data is obtained, why some values are being added or modified, and  which experiments or analyses are being performed to obtain a set of values. 
Annotations can be also used to track the provenance of the data, e.g., from which source a piece of data is obtained or which program is used to generate the data. 
Tracking the provenance of the data is very important in assessing the value and credibility of the data
and in giving credit to the original data generators. 

Users can annotate the data at multiple granularities, e.g., annotating an entire table, an entire column, 
a subset of the tuples, a few cells, or a combination of these.

Despite their importance, annotations are not systematically supported by most database systems.  
While annotation management has been addressed in previous works,
e.g.,~\cite{ref62, ref63, ref60, ref61},
most of the proposed techniques usually assume simple annotation schemes
and focus mainly on annotation propagation, i.e., propagating annotations along with the query answers.
Other aspects of annotations management, e.g., mechanisms for their insertion, archival, and indexing 
as well as more efficient annotation schemes such as multi-granularity schemes, have not been addressed.

In bdbms, we address several challenges and requirements of annotation management. 
We highlight these challenges and requirements through the following example. 
We consider two gene tables, {\tt DB1\_Gene} and {\tt DB2\_Gene} that have been obtained from two different databases (Refer to Figures~\ref{fig:ann-example1} and~\ref{fig:ann-example2} for illustration). 
Each table has a set of annotations attached to it. 
We assume a straightforward storage scheme for storing the annotations, e.g.,  the one used in~\cite{ref62}, 
where each column in the database has an associated annotation column to store the annotations  
(Figure~\ref{fig:ann-example2}). 

{\textbf{Adding annotations:}}
Users want to annotate their data at various granularities in a transparent way. That is, how or where the annotations are stored should be transparent from end-users. 
However, current database systems do not provide a mechanism to facilitate annotating the data. 
For example, to add annotation {\em A2} over table {\tt DB1\_Gene}  (Figure~\ref{fig:ann-example1}), the user has to know that the annotations 
are stored in the same user table in columns {\em Ann\_GID}, {\em Ann\_GName}, and 
{\em Ann\_GSequence}. Then, the user issues an UPDATE statement to update these columns by adding {\em A2} 
to the desired annotation cells (Figure~\ref{fig:ann-example2}). 

To support data annotation, the system has to provide new mechanisms for seamlessly adding annotations at various granularities.
It is essential to provide new expressive commands as well as visualization tools
that allow users to add their annotations graphically.
 
{\textbf{Storing annotations at multiple granularities:}}
As in Figure~\ref{fig:ann-example1}, users may annotate a single cell, e.g., {\em A3}, few cells, e.g., {\em A1 and B2}, entire rows, e.g., {\em A2 and B4}, 
or entire columns, e.g., {\em B3}. Multi-granularity annotations motivate the need for efficient storage and indexing schemes.
Otherwise, storing and processing the annotations can be very expensive. 
For example, annotations {\em A2} and {\em B3} are repeated in the annotation columns 6 and 5 times, respectively.    
The need for such efficient schemes is especially important in the context of provenance
where a single provenance record can be attached to
many tuples or even entire columns or tables.

{\textbf{Categorizing  annotations:}}
Although all annotations are metadata, they may have different importance, meaning, and creditability. 
For example, annotations that are added by a certain user or group of users can be more important than annotations added by the public or unknown users. 
Moreover, annotations that represent the lineage of the data have different purpose and importance from the annotations that represent users' comments. 
For example, annotations {\em A2} and {\em B3} represent the lineage of some data, and users may be interested only in these annotations. 
As will be discussed later in the paper, the different types of annotations will also have an impact on the storage mechanism adopted for each type. 
This diversity in annotations motivates the need for separating or categorizing the annotations. 
bdbms provides a mechanism that allows users to categorize their annotations at the storage, query processing, and annotation propagation levels.

{\textbf{Archiving annotations:}} 
Users may need to archive or delete annotations as they become obsolete, old, or simply invalid. 
Archived annotations should not be propagated to users along with query answers.   
For example, annotation {\em B5} in Figure~\ref{fig:ann-example1} states that gene {\em JW0080} has an unknown function. But 
if the function of this gene becomes known and gets added to the database, then {\em B5} becomes invalid and users do not want to propagate this annotation along 
with query answers. 
Without providing a mechanism for archiving annotations, the archival operation may not be an easy task. 
For example, to archive annotation {\em B5}, the user needs to find out which tuples/cells in the database has {\em B5}, then the contents of 
each of these cells are parsed to archive then delete {\em B5}.

{\textbf{Propagating annotations:}}
A key requirement in allowing annotation propagation is to simplify users' queries.
This can be only achieved by
providing database system support for annotation propagation; for
example, by extending the  query operators. Otherwise, users' queries may become complex and user-unfriendly.
For example, consider a simple query that retrieves the genes that are common in {\tt DB1\_Gene} and {\tt DB2\_Gene} along with their 
annotations (Figure~\ref{fig:ann-example2}). 
To answer this query, the user has to write the following SELECT statements (a--c):\\ 
 
$(a)~~R_1 (GID, GName, GSequence)$ =\\ 
{ {\em
$~~~~~~~~~~~~~$SELECT  GID, GName, GSequence\\ 
$~~~~~~~~~~~~~$FROM   DB1\_Gene\\
$~~~~~~~~~~~~~$INTERSECT\\
$~~~~~~~~~~~~~$SELECT  GID, GName, GSequence\\ 
$~~~~~~~~~~~~~$FROM   DB2\_Gene;\\
}}

In Step (a), the user selects only the data columns from both gene tables, i.e., {\em GID, GName, GSequence}, and performs the intersection operation.\\  

$(b)~~R_2 (GID, GName, GSequence, Ann\_GID,$\\
$~~~~~~~~~~~~Ann\_GName, Ann\_GSequence)=$\\
{ {\em
$~~~~~~~~~~~~~$SELECT  R.GID, R.GName, R.GSequence,\\ 
$~~~~~~~~~~~~~~~~~$G.Ann\_GID, G.Ann\_GName, G.Ann\_GSequence\\
$~~~~~~~~~~~~~$FROM   R\_1 R, DB1\_Gene G\\
$~~~~~~~~~~~~~$WHERE  R.GID = G.GID;\\
}}

In Step (b), the user joins the output from Step (a) back with Table {\tt DB1\_Gene} in order to retrieve the annotations from this table.
Notice that we cannot select the annotation columns in Step (a) because, since the annotation values in the annotation attributes may vary in the two tables, in this case the intersection operation would not return any tuples.  \\

$(c)~~R_3 (GID, GName, GSequence, Ann\_GID,$\\
$~~~~~~~~~~~~Ann\_GName, Ann\_GSequence) =$\\
{ {\em
$~~~~~~~~~~~~~$SELECT  R.GID, R.GName, R.GSequence,\\
$~~~~~~~~~~~~~~~~~$R.Ann\_GID+G.Ann\_GID,\\
$~~~~~~~~~~~~~~~~~$R.Ann\_GName+G.Ann\_GName,\\
$~~~~~~~~~~~~~~~~~$R.Ann\_GSequence+G.Ann\_GSequence\\
$~~~~~~~~~~~~~$FROM   R\_2 R, DB2\_Gene G\\
$~~~~~~~~~~~~~$WHERE  R.GID = G.GID;\\
}}

In Step (c), a join is performed between $R_2$ and {\tt DB2\_Gene}  to consolidate the annotations from 
{\tt DB2\_Gene} with $R_2$'s annotations, where $+$ is the annotation union operator.

The main reason for the complexity of querying and propagating the annotations is that users view annotations as metadata, 
whereas the DBMSs view annotations as normal data. 
For example, from a user's view point, the two tuples corresponding to genes {\em JW0080} and {\em JW0055} in Table {\em DB1\_Gene} 
are identical to those in Table {\em DB2\_Gene} (Figure~\ref{fig:ann-example2}). They only have different annotations. 
Whereas, from the database view point, these tuples are not identical because annotations are viewed as normal attribute data. 
As a result, users' queries may become complex in order to overcome the mismatch in interpreting the annotations. 

In the following subsections, we introduce our initial investigations through bdbms to address the challenges and requirements highlighted above along 
with some preliminary results. 

\begin{figure}[t]
 \centering
   \includegraphics[height=2cm, width= 7cm, angle=0]{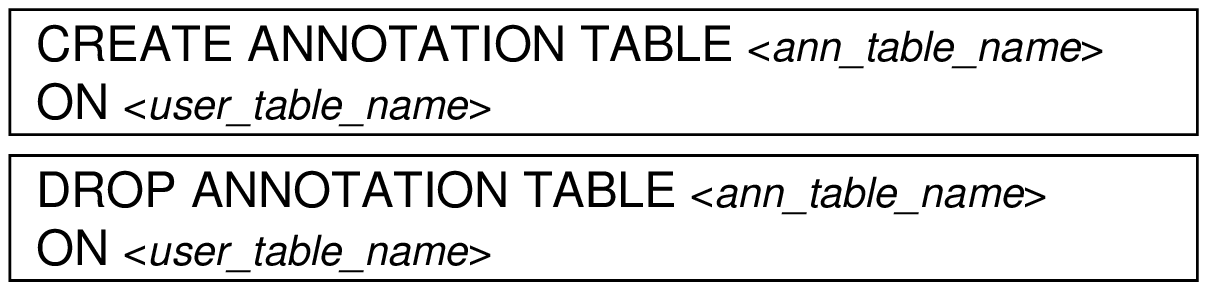}
      \caption{The A-SQL commands CREATE and DROP}
    \label{fig:create}
\end{figure}

\begin{figure}[t]
 \centering
   \includegraphics[height=5cm,width=8cm]{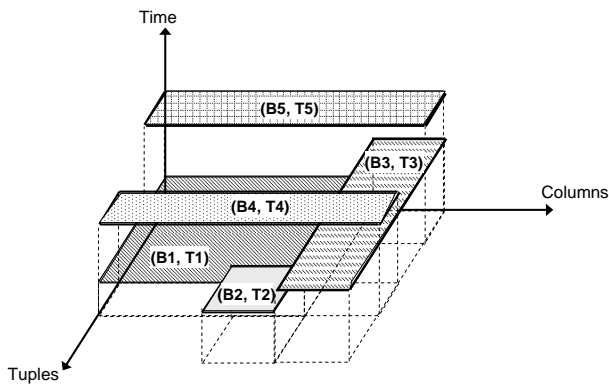}
      \caption{Compact storage for annotations}
    \label{fig:index}
\end{figure}

\subsection{Storing and Indexing Annotations}  
bdbms allows a user relation to have multiple annotation tables attached to it. 
For example, table {\tt DB1\_Gene} may have an annotation table that stores the provenance information and another annotation table that stores users' comments.  
To create an annotation table over a given user relation, the A-SQL command {\em CREATE ANNOTATION TABLE} (Figure~\ref{fig:create}) is used. 
{\em CREATE ANNOTATION TABLE} allows users to design and categorize their annotations at the storage level. 
This categorization will also facilitate annotation propagation (discussed in  Section~\ref{propagate-anno}), where users may request propagating a certain type of annotations. 
To drop an annotation table, the {\em DROP ANNOTATION TABLE} command (Figure~\ref{fig:create}) is used.

To efficiently store the annotations, we are investigating
several storage and indexing schemes. 
One possible direction is to consider compact representation of annotations 
that would improve the system performance with respect to storage overhead, I/O cost to retrieve the annotations, 
and the query processing time. 
For example, instead of storing the annotations at the cell level, we may store some of the annotations at coarser granularities. 
For instance, the annotations over Table {\tt DB2\_Gene} (Figure~\ref{fig:ann-example1}) 
can be represented as rectangles attached to groups of contiguous cells 
as illustrated in Figure~\ref{fig:index}, where {\tt DB2\_Gene} is viewed as two-dimensional space, e.g.,  
columns represent the X-axis and tuples represent the Y-axis.  
In this case, an annotation over any group of contiguous cells can be represented by a single annotation record. So, in general, an annotation over a subset of a table will map to multiple rectangular regions. 
Other annotation characteristics that may need to be taken into account 
include whether the annotation  is linked to multiple
data items in different tables or is linked to very few specific cells.

\subsection{Adding Annotations at Multiple Granularities}  
\label{add-anno}

\begin{figure}[t]
 \centering
   \includegraphics[height=4cm, width= 8cm, angle=0]{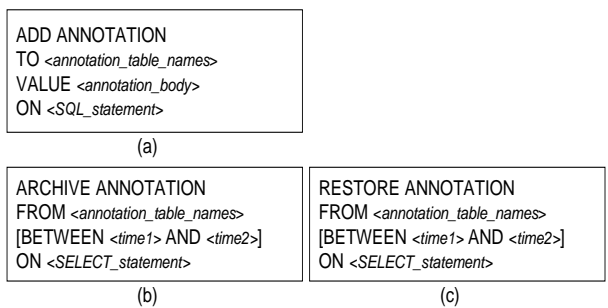}
      \caption{The A-SQL commands ADD, ARCHIVE, and RESTORE}
    \label{fig:add}
\end{figure}

To add annotations using A-SQL, we propose the {\em ADD ANNOTATION} command (Figure~\ref{fig:add}(a)). 
The {\em annotation\_table\_names} specifies to which annotation table(s) the 
added annotation will be stored. 
The {\em annotation\_body} specifies the annotation value to be added. 
The output of the {\em SQL\_statement} specifies the data to which the annotation is attached. 
Since annotations may contain important information that users want to query, we plan to support XML-formatted annotations. 
That is, {\em annotation\_body} is an XML-formatted  text. In this case, users can (semi-)structure their annotations and 
make use of XML querying capabilities over the annotations. 
The output of the {\em SQL\_statement} can be at various granularities, e.g., entire tuples, columns, or group of cells. 
For example, to add annotation {\em B3} over the entire {\em GSequence} column in Table {\tt DB2\_Gene} (as illustrated in Figure~\ref{fig:ann-example1}), 
we execute the following ADD ANNOTATION command: 

{\small{\em 
$~~~~~$ADD ANNOTATION\\ 
$~~~~~~~~~$TO DB2\_Gene.GAnnotation\\ 
$~~~~~~~~~$VALUE '$<Annotation>$\\ 
$~~~~~~~~~~~~~~~~~~~~~~~~~$obtained from GenoBase \\ 
$~~~~~~~~~~~~~~~~~~~~</Annotation>$'\\ 
$~~~~~~~~~$ON (Select  G.GSequence\\ 
$~~~~~~~~~~~~~~~$From    DB2\_Gene G);\\ 
}} 

In this case, the annotation is attached to the entire {\em GSequence} column because no WHERE clause is specified. 
The annotation is stored in the annotation table {\em GAnnotation}. 
Notice that $<Annotation>$ is the XML tag that encloses the annotation information. 

Similarly, to annotate an entire tuple, e.g., annotation {\em B5}, 
we execute the following ADD ANNOTATION command:

{\small{\em 
$~~~~~$ADD ANNOTATION\\ 
$~~~~~~~~~$TO DB2\_Gene.GAnnotation\\ 
$~~~~~~~~~$VALUE '$<Annotation>$\\ 
$~~~~~~~~~~~~~~~~~~~~~~~~~$This gene has an unknown function \\ 
$~~~~~~~~~~~~~~~~~~~~</Annotation>$'\\ 
$~~~~~~~~~$ON (Select  G.*\\ 
$~~~~~~~~~~~~~~~$From    DB2\_Gene G\\ 
$~~~~~~~~~~~~~~~$WHERE  GID = 'JW0080');\\ 
}} 

In this case, the annotation is attached to the entire 
tuples returned by the query  since all the attributes in the table are selected.

To allow users to link annotations to database operations, i.e.,  INSERT, UPDATE, or DELETE, the {\em SQL\_statement} 
will be an INSERT, UPDATE or DELETE statement. 
For example, instead of inserting a new tuple and then annotating it by issuing a separate ADD ANNOTATION command, 
users can insert and  annotate the new tuple instantly by enclosing the insert statement inside the ADD ANNOTATION command. 
For the delete operation, the deleted tuples will be stored in separate log tables along with the annotation that specifies why these tuples 
have been deleted.  Notice that the standard system recovery log cannot be used for this purpose as the users need the freedom to structure their annotation schemas the way they want, which system recovery logs do not support.

We plan to add a visualization tool to allow users to annotate their data in a transparent way.
The visualization tool displays users' tables as grids or spreadsheets where users can select one or more cells to annotate.
Oracle address the integration of  database tables with Excel spreadsheets
to make use of Excel visualization and analysis power~\cite{p22}.
In bdbms, we plan to add this integration feature to facilitate adding and visualizing annotations.

\subsection{Archiving and Restoring Annotations}  
\label{archive-anno}
Archival of annotations allows users to isolate old or invalid annotations from recent and valuable ones. 
In bdbms, we support archival of annotations instead of  permanently deleting them because biological data usually has a degree of uncertainty 
and old values may turn out to be the correct values. Archiving annotations gives users the flexibility to restore the annotations back if needed.   
Unlike other annotations, archived annotations are not propagated to users along with the query answers. However, 
if archived annotations are restored, then they will be propagated normally.  

To archive and restore annotations, we introduce the {\em ARCHIVE ANNOTATION} (Figure~\ref{fig:add}(b)) and 
{\em RESTORE ANNOTATION} (Figure~\ref{fig:add}(c)) commands, respectively. 
The FROM clause specifies from which annotation table(s) the annotations will be archived/restored. 
The optional clause BETWEEN specifies a time range over which the annotations will be archived/restored. 
This time corresponds to the times-tamp assigned to each annotation when it is first added to the database.
The output from the {\em SELECT\_statement} specifies the data on which the annotations will be archived/restored. 
In addition, the output from the {\em SELECT\_statement} can be at multiple granularities, as explained in the {\em ADD ANNOTATION} command.

\subsection{Annotation Propagation and Annotation-based Querying}
\label{propagate-anno}
\begin{figure}[t]
 \centering
   \includegraphics[height=3.7cm, width= 7cm, angle=0]{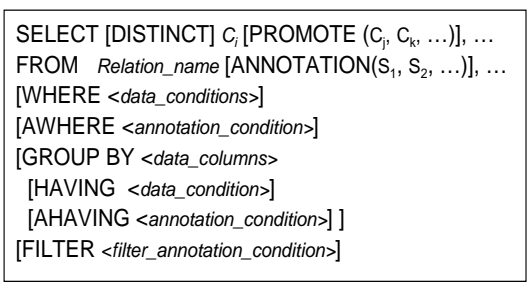}
      \caption{The A-SQL SELECT command}
    \label{fig:aselect}
\end{figure}

To support the propagation of annotations and querying of the data based on their annotations, 
we introduce the A-SQL command SELECT, given in Figure~\ref{fig:aselect}.
A-SQL SELECT extends the standard SELECT by introducing new operators and extending the semantics of the standard operators. 
We introduce the new operators ANNOTATION, PROMOTE, AWHERE, AHAVING, and FILTER. 

The ANNOTATION operator allows users to specify which annotation table(s) to consider in the query. 
Using the ANNOTATION operator, users can propagate their annotations transparently. That is, users do not have to know how or where annotations are stored. Instead, users only specify which annotations are of interest. 

The PROMOTE operator allows users to copy annotations from one or more columns, possibly not in the projection list, to a projected column.  
For example, if column {\em GID} is projected from Table {\tt DB1\_Gene}, then Annotation {\em A3} will not be propagated unless the annotations over 
{\em GSequence} are copied to {\em GID}.  

The AWHERE and AHAVING clauses are analogous to the standard WHERE and HAVING clauses except that the conditions of 
AWHERE and AHAVING are applied over the annotations. 
That is, AWHERE and AHAVING pass a tuple along with all its annotations only if the tuple's annotations satisfy the given AWHERE and AHAVING conditions.   
On the other hand, the FILTER clause passes all the data tuples of the input relation (keeps user's data intact) but it filters the annotations attached to each tuple. 
That is, any annotation that does not satisfy {\em filter\_annotation\_condition} is dropped.  
 
The standard operators, e.g., {\em projection}, {\em selection}, and {\em duplicate elimination}, are also extended to process the annotations attached 
to the tuples. 
For example, the {\em projection} operator selects some user attributes from the input relation 
and passes only the annotations attached to those attributes. 
For example, projecting column {\em GID} from Table {\tt DB2\_Gene} (Figure~\ref{fig:ann-example1}) 
results in reporting {\em GID} data along with annotations {\em B1, B4,} and {\em B5} only. 
The {\em selection} operators in WHERE and HAVING select tuples from the input relation based on conditions applied over the data values. 
The selected tuples are passed along with all their annotations. 
For example, selecting the gene with {\em GID} $=$ {\em JW0080} from Table {\tt DB2\_Gene} results in reporting the 
first tuple in {\tt DB2\_Gene} along with annotations {\em B1, B3,} and {\em B5}.       
Operators that group or combine multiple tuples into one tuple, e.g., {\em duplicate elimination}, {\em group by}, {\em union}, {\em intersect}, 
and {\em difference}, are also extended to handle the annotations attached to the tuples. 
These operators union the annotations over the grouped or combined tuples and attach them to the output tuple that represents the group.

While defining the above commands and operators is only the first step in supporting annotations and other features
within bdbms,
we need to define for each
A-SQL operator its algebraic definition, cost estimate function, and algebraic properties
that can be used by the query optimizer to generate efficient query plans.

\end{sloppypar}

\section{Provenance Management}
\label{sec:provenance}

\begin{figure}
 \centering
   \includegraphics[height=5cm, width= 8.5cm, angle=0]{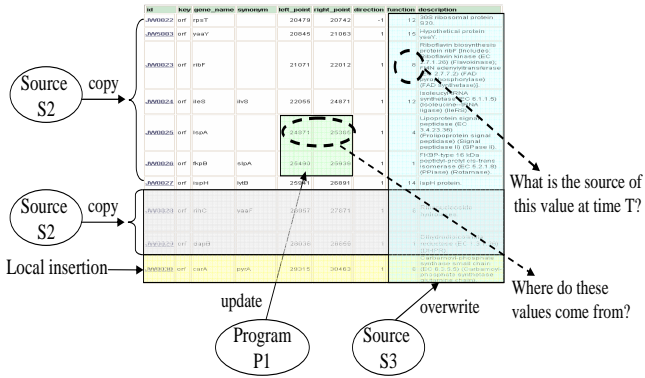}
      \caption{Data provenance at multiple granularities}
    \label{fig:provenance}
\end{figure}

Biologists commonly interact and exchange data with each other. 
Tracking the provenance (lineage) of data is very important in assessing the value and credibility of the data. 
Similar to annotations, data provenance can be attached to the database at multiple granularities, 
i.e., at the table, column, tuple levels, or any sub-groupings and subsets of the data. Also, biological data can be queried by its provenance. 
For example (refer to Figure~\ref{fig:provenance}), one table may contain data from multiple sources, e.g., {\em S1} and {\em S2}, or data that is locally inserted. 
Then, some values may be updated by a certain program, e.g., {\em P1}, and some columns may be overwritten by data from another source, e.g., {\em S3}.  
Then, users may be interested to know the source of some values at a certain moment in time. 

\begin{sloppypar}
In bdbms, we treat provenance data as a kind of annotations where all the requirements and functionalities discussed in Section~\ref{sec:annotation} 
are also applicable to provenance data. 
However, provenance data has special requirements and characteristics that need to be addressed including:
\begin{itemize}
\item \textbf{Structure of provenance data:} 
Unlike annotations that can be free text, provenance data usually has well-defined structure. 
For example, the names of the source database and the source table draw their values from a list of pre-defined values. 
Supporting XML-formatted annotations can be beneficial in structuring provenance data. 
For example, provenance data can follow a predefined XML schema that needs to be stored and enforced by the database system. 

\item \textbf{Authorization over provenance data:} 
End-users are usually not allowed to insert or update the provenance data. Provenance data needs to be automatically inserted and maintained by the system. 
For example, integration tools that copy the data from one database to another can be the only tools that insert the provenance information. 
End-users can only retrieve or propagate this information.
Therefore, we need to provide an access control mechanism over the provenance data (and annotations in general) to restrict 
the annotation operations, e.g., addition, archival, and propagation, to certain users or programs as required.   
\end{itemize}
\end{sloppypar}


\begin{figure}
 \centering
   \includegraphics[height=7.5cm,width=8.5cm]{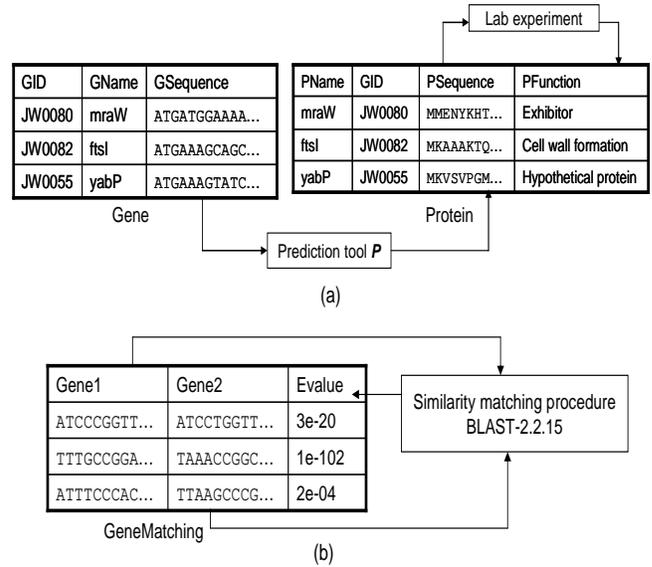}
      \caption{Local dependency tracking}
    \label{fig:dependency3}
\end{figure}

\section{Local Dependency Tracking}
\label{sec:dependency}
\begin{sloppypar}
Biological databases are full of dependencies and derivations among data items. 
In many cases, these dependencies and derivations cannot be automatically computed using coded functions, e.g., stored procedures or functions inside the database. 
Instead, they may involve prediction tools, lab experiments, or instruments to derive the data. 
Using integrity constraints and triggers to maintain the consistency of the data is limited to computable dependencies, 
i.e., dependencies that can be computed via coded functions. 
However, non-computable dependencies cannot be directly handled using integrity constraints and triggers. 
In Figure~\ref{fig:dependency3}, we give an example of the dependencies that can be found in biological databases. 
In Figure~\ref{fig:dependency3}(a), protein sequences are derived from the gene sequences using  
a prediction tool {\em P}, whereas the function of the protein is derived from the protein sequence using lab experiments.  
If a gene sequence is modified, then all protein sequences that depend on that gene have to be marked as {\em outdated} until their values  are re-verified. 
Moreover, the function of the outdated proteins has to be marked as {\em outdated}  until their values are re-evaluated.  

In Figure~\ref{fig:dependency3}(b), we present another type of dependency where the value of the data in the database 
depends on the procedure or program that generated that data. 
For example, the values in the {\em Evalue} column  (Figure~\ref{fig:dependency3}(b)) depend on Procedure {\em BLAST-2.2.15}. 
If a newer version of {\em BLAST} is used or {\em BLAST} is replaced with another procedure, then we need to re-evaluate the values in the {\em Evalue} column. 
These values can be automatically evaluated if {\em BLAST} can be modeled as a database function. Otherwise, the values have to be marked as {\em Outdated}.

In bdbms, we propose to extend the concept of {\em Functional Dependencies}~\cite{p8, p9} to {\em Procedural Dependencies}. 
In {\em Procedural Dependencies}, we not only track the dependency among the data, but also the type and characteristics  of the dependency, e.g.,
the procedure on which the dependency is based, whether or not that procedure can be executed by the database, and 
whether or not that procedure is invertible. 
For example, we can model the dependencies in Figure~\ref{fig:dependency3} using the following rules.  

\begin{figure}[h]
 \centering
   \includegraphics[height=4cm,width=8cm]{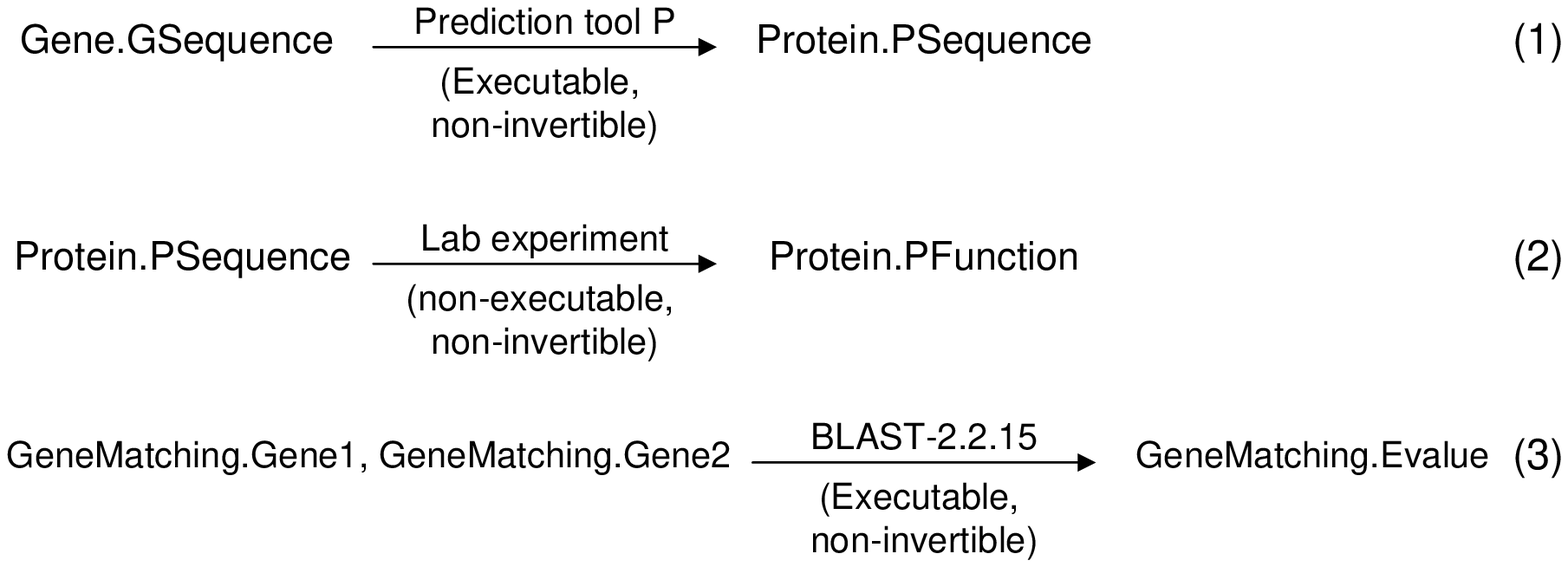}
    \label{fig:rules}
\end{figure}

Rule 1 specifies that Column {\em PSequence} in Table {\tt Protein} depends on Column {\em GSequence} in Table {\tt Gene} through the prediction tool {\em P} that 
is executable by the database and is non-invertible. 
Rule 2 specifies that column {\em PFunction} in Table {\tt Protein} 
depends on Column {\em PSequence} through a lab experiment that is not executable by the database and is non-invertible. 
Rule 3 specifies that Column {\em Evalue} in Table {\tt GeneMatching} depends on both columns {\em Gene1} and {\em Gene2} through Program {\em BLAST-2.2.15} 
that is executable by the database and is non-invertible. 
For example, from Rule 2, we infer that  when Column {\em PSequence} changes, the database can only mark {\em PFucntion} as {\em Outdated}. In contrast, based on Rule 3, when either of the {\em Gene1} or {\em Gene2} columns or Procedure {\em BLAST-2.2.15} change, the   
database can automatically re-evaluate {\em Evalue}.

In addition, the notion of {\em Procedural Dependencies} allows us to reason about the dependency rules. 
For example, in addition to the closure of an attribute, we can compute the closure of a procedure, i.e., all data in the database that depend on a specific procedure. 
We can also derive new rules, for example, based on rules (1) and (2) above, we can derive the following rule:

\begin{figure}[h]
 \centering
   \includegraphics[height=1cm,width=8cm]{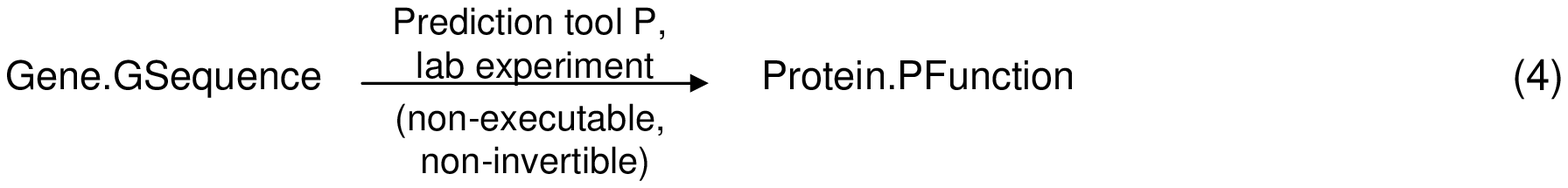}
    \label{fig:rules2}
\end{figure}

Rule 4 specifies that Column {\em PFunction} in Table {\tt Protein} depends on Column {\em GSequence} in Table {\tt Gene} through a chain of two procedures, 
a perdition tool {\em P} and a lab experiment. This chain is non-executable by the database and is non-invertible. 
Notice that the chain is  non-executable because at least one of the procedures, namely the lab experiment, is non-executable. 
 
\begin{figure}[t]
 \centering
   \includegraphics[height=2.7cm,width=8cm]{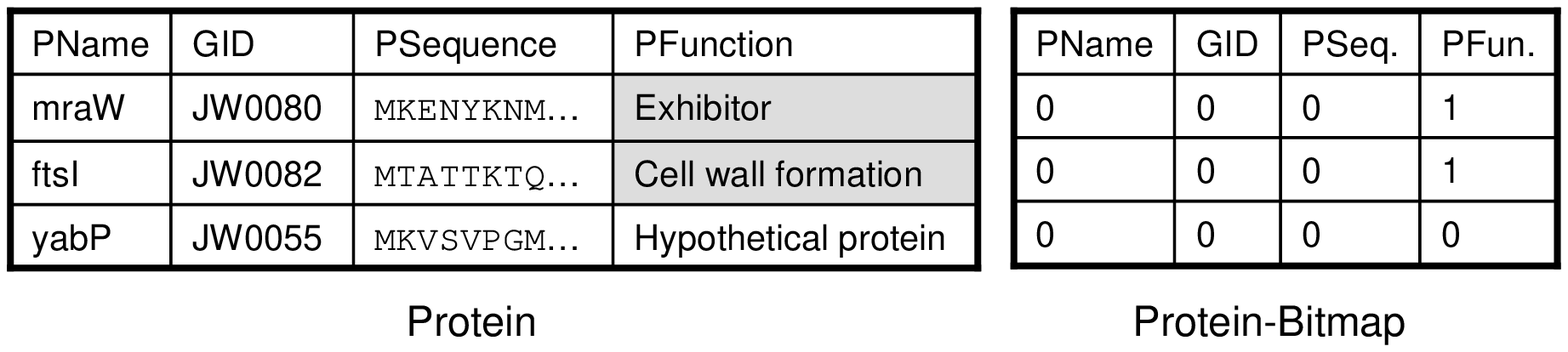}
   \caption{Use of bitmaps to mark outdated data}
    \label{fig:bitmap}
\end{figure}

In bdbms, we address the following functionalities to track local dependencies:
\begin{itemize}

\item { \textbf{Modeling dependencies:} 
We use {\em Procedural Dependencies} to allow users to model the dependencies among the database items as well as for bdbms to reason about these dependencies, for example, 
to detect conflicts and cycles among dependency rules, and to compute the closure of procedures. 
}

\item { \textbf{Storing dependencies:} 
Dependencies among the data can be either at the schema level, 
i.e., the entity level, or at the instance level, i.e., the cell level. Schema-level dependencies can be modeled using foreign key constraints, 
e.g., protein sequences depend on gene sequences and they are linked by a foreign key. Instance-level dependencies are more complex to model because they are on a 
cell-by-cell basis.  In this case, we can use dependency graphs to model such dependencies. 
}

\item { \textbf{Tracking outdated data:} 
When the database is modified, bdbms uses the dependency graphs to figure out which items, termed the {\em outdated} items, may be affected by this modification. 
{\em Outdated} items need to be marked such that these items can be identified in any future reference. 
We propose to associate a bitmap with 
each table in the database. A cell in the bitmap is set to 1 if the corresponding cell 
in the data table is {\em outdated}, otherwise the bitmap cell is set to 0. 
For example, assume that the sequences corresponding to genes {\em JW0080} and {\em JW0082} (Figure~\ref{fig:dependency3}(a)) are modified, 
then the bitmap associated with Table {\tt Protein} will be as illustrated in Figure~\ref{fig:bitmap}. 
Notice that the bits corresponding to {\em PSequence} are not set to 1 because {\em PSequence} is automatically updated by executing Procedure {\em P}. 
In contrast, {\em PFunction} cannot be automatically updated, therefore its corresponding bits are set to 1 to indicate that these values are outdated.  
To reduce the storage overhead of the maintained bitmaps, data compression techniques such as Run-Length-Encoding~\cite{p10} 
can be used to effectively compress the bitmaps. 
}

\item { \textbf{Reporting and annotating outdated data:} 
The main objective of tracking local dependencies is that the database should be able to report  at all times 
the items that need to be verified or re-evaluated.  
Moreover, when a query executes over the database and involves {\em outdated} items, 
the database should propagate with those items an annotation specifying 
that the query answer may not be correct. 
Detecting the {\em outdated} items at query execution time is a challenging problem as it requires retrieving and propagating 
the status of each item, i.e., whether it is outdated or not, in the query pipeline. A proposed solution is to consider the status of the database items as 
annotations attached to those items. These annotations will be automatically propagated along with the query answers as discussed in Section~\ref{sec:annotation}.  
}

\item {\textbf{ Validating outdated data:}
bdbms will provide a mechanism for users to validate {\em outdated} items. 
An {\em outdated} item may or may not need to be modified to become valid. 
For example, a modification to a gene sequence may not affect the corresponding protein sequence. In this case, the protein sequence will be revalidated without 
modifying its value. 
}
\end{itemize}

\end{sloppypar}


\section{Update Authorization}
\label{sec:approval}

\begin{sloppypar}
Changes over the database may have important consequences, and hence, 
they should be subject to authorization and approval by authorized entities before these changes become permanent in the database. 
Update authorization (also termed {\em approval enforcement}) in current database management systems is based on GRANT/REVOKE access models~\cite{ref71, ref70}, 
where a user may be granted an authorization to update a certain table or attribute. 
Although widely acceptable, these authorization models are based only on the identity of 
the user not on the content of the data being inserted or updated. 
In biological databases, it is often the case that a data item can make it permanently to a database based on its value not on
the user who entered that value.
For example, a lab administrator may allow his/her lab members
to perform insert and update operations over the database.
However, for reliability, these operations have to be revised by the lab administrator.
If the lab administrator is the only user who has the right to update the database, then this person may become
a bottleneck in the process of populating the database.
\end{sloppypar}

\begin{figure}[t]
 \centering
   \includegraphics[height=2cm, width= 8cm, angle=0]{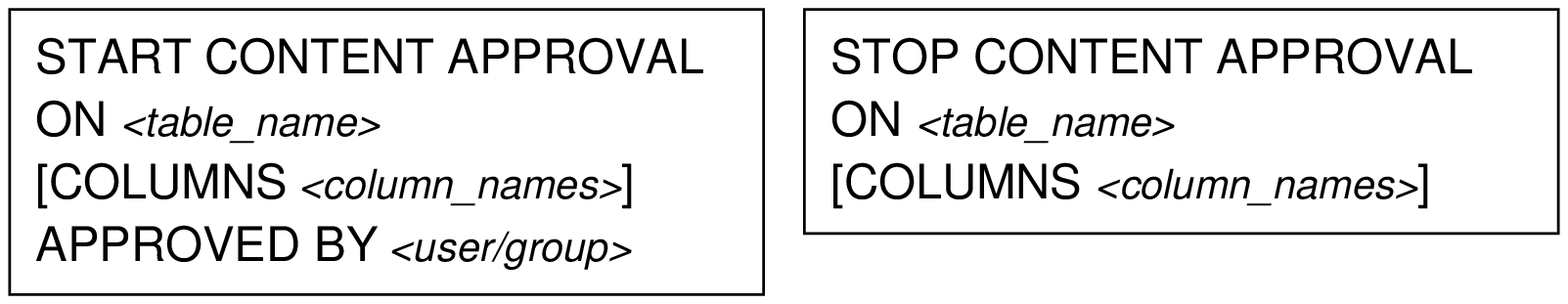}
      \caption{Content-based approval}
    \label{fig:approval}
\end{figure}

\begin{sloppypar}
In bdbms, we introduce an approval mechanism, termed {\em content-based approval}, that allows the database to systematically track the changes 
over the database. 
The proposed content-based approval mechanism works with, not in replacement
to, existing GRANT/REVOKE mechanisms.
The content-based approval mechanism maintains a log of all update operations,
 i.e., INSERT, UPDATE, and DELETE, 
that occur in the database. 
The database administrator can turn the 
content-based approval feature ON or OFF for a certain table or columns  
using a {\em Start Content Approval} and  {\em End Content Approval}  commands (Figure~\ref{fig:approval}), respectively. 
The {\em table\_name} value specifies the user table on which the update operations will be monitored. 
The optional clause COLUMNS specifies which column(s) in {\em table\_name} to monitor. For example, we can monitor the update operations over only Column {\em GSequence}
of Table {\tt Gene} (Figure~\ref{fig:dependency3}(a)). The APPROVED BY clause specifies the user or group of users who can approve or disapprove the update operations.  
If the content-based approval feature is turned ON over Table T, then bdbms stores all update operations over T in the log along with an automatically generated 
{\em inverse} statement that negates the effect of the original statement.  
More specifically, for INSERT, a DELETE statement will be generated, for DELETE, an INSERT statement will be generated, 
and for UPDATE, another UPDATE statement that restores the old values will be generated. 
The log stores also the user identifier who issued the update operation and the issuing time. 
The person in charge of the database, e.g., the lab administrator, can then view the maintained log and revise 
the updates that occurred in the database.
If an operation is disapproved, then bdbms executes the inverse statement of that operation to remove its effect from the database. 
Executing the inverse statement may affect other elements in the database, e.g., elements that depend on the currently existing values. 
It is the functionality of the {\em Local Dependency Tracking} feature (Section~\ref{sec:dependency}) to 
track and invalidate these elements. 
\end{sloppypar}

\section{Indexing and Query Processing}
\label{sec:indexing}
\begin{sloppypar}
Biological databases warrant the use of non-traditional indexing mechanisms beyond B+-trees
and hash tables. To enable biological algorithms to operate efficiently on the database, we propose  
integrating non-traditional indexing techniques inside bdbms. We focus on two fronts: 
(1) Supporting multidimensional datasets via multidimensional indexing techniques (suitable
for protein 3D structures and surface shape matching), and (2) Supporting compressed datasets via novel 
external-memory indexes that work over the compressed data without decompressing it (suitable for indexing large sequences). 

In bdbms, we focus on introducing non-traditional index structures for supporting biological data. 
For example, compressing the data inside the database is proven to improve the system performance, e.g., C-store~\cite{ref58}. 
It reduces significantly the size of the data, the number of I/O operations required to retrieve the data, and the buffer requirements. 
In bdbms, we investigate how we can store biological data in compressed form and yet be able to 
operate, e.g., index, search, and retrieve, on the compressed data without decompressing it.

\subsection{Indexing Multi-dimensional Data}
\label{sec:sp-gist}
Space-partitioning trees are a family of access methods that index objects in a multi-dimensional space, e.g., protein 3D structures. 
In~\cite{p2, p1, p4, p3}, we introduce an extensible indexing framework, termed SP-GiST, that broadens the class of supported indexes to
include disk-based versions of space-partitioning trees, e.g., disk-based
trie variants, quadtree variants, and kd-trees. 
As an extensible indexing framework, SP-GiST allows developers to instantiate a variety of index structures in an efficient way through pluggable modules 
and without modifying the database engine. 
The SP-GiST framework is implemented inside PostgreSQL~\cite{p19} and we use it in bdbms. 
Several index structures have been instantiated using SP-GiST, e.g., variants of the trie~\cite{p11, p12}, the kd-tree~\cite{p13}, 
the point quadtree~\cite{p15}, and the PMR quadtree~\cite{p14}. 
We implemented several advanced search operations, e.g., k-nearest-neighbor search, 
regular expression match search, and substring searching.  
The experimental results in~\cite{p4} demonstrate the performance potential of the class of 
space-partitioning tree indexes over the B+-tree and R-tree indexes, for the operations above. 
In addition to the performance gains and the advanced search 
functionalities provided by SP-GiST indexes, it is the ability to rapidly prototype these 
indexes inside bdbms that is most attractive.  

A key challenge is to integrate SP-GiST indexes inside biological analysis algorithms 
such as protein structure alignment algorithms. 
Providing the index structures is the first step to improve the querying and processing capabilities of the analysis algorithms.

\subsection{Indexing Compressed Data}
\label{sec:compressed}
\begin{figure}
 \centering
   \includegraphics[height=7.5cm, width= 8cm]{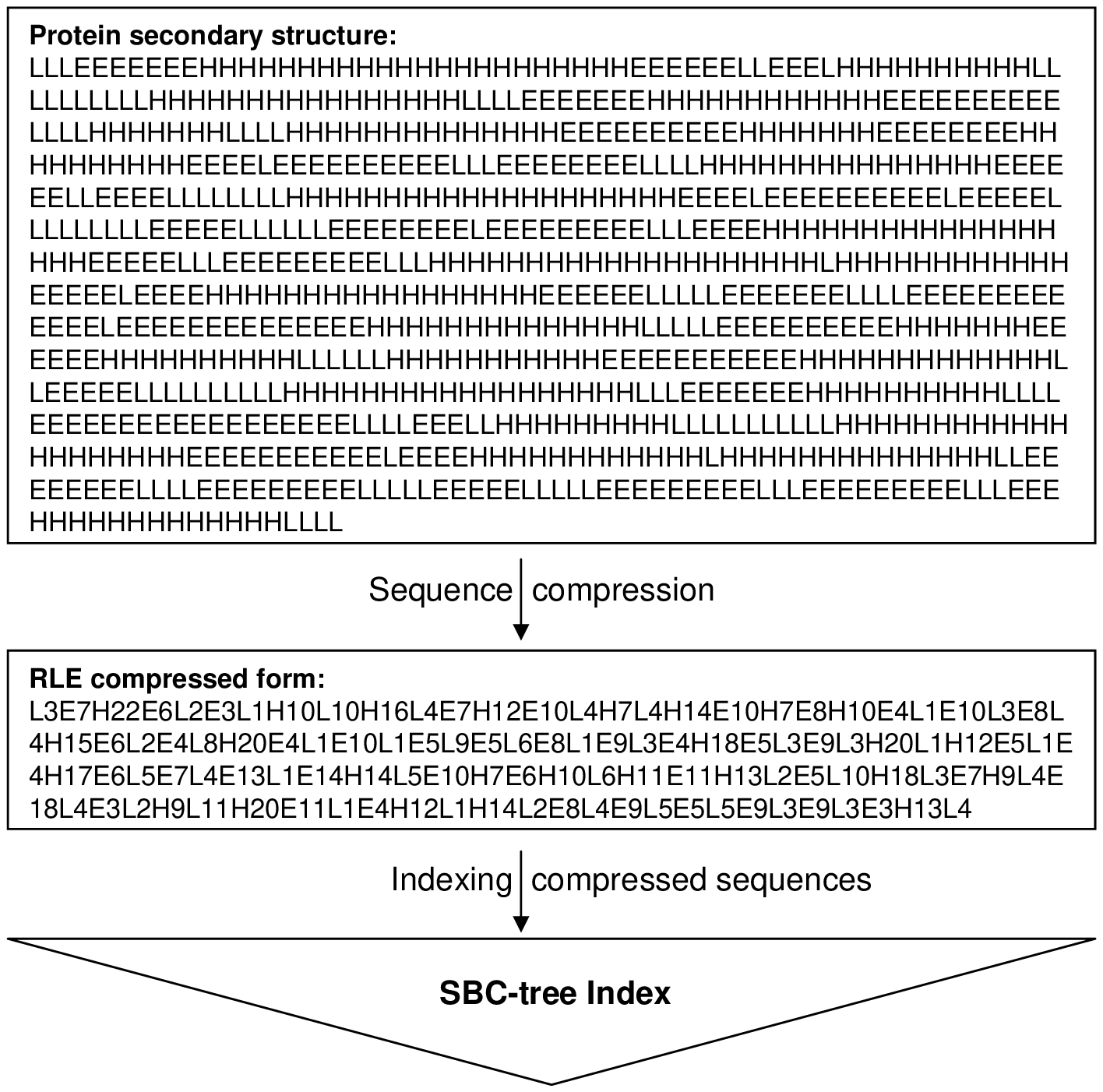}
      \caption{Indexing and querying RLE-compressed sequences}
    \label{fig:compression}
\end{figure}
Biological databases consist of large amounts of sequence data, e.g., genes, alleles, and protein primary and secondary structures. 
These sequences need to be stored, indexed, and searched efficiently. 
In bdbms, we propose to investigate new techniques
for compressing biological sequences and operating over the compressed data without decompressing it.
Sequence compression has been addressed recently in the C-Store database management system~\cite{ref58}, where some operators, e.g., 
aggregation operators, can operate directly
over the compressed data. Sequence compression is demonstrated to improve system performance as it reduces the size of the data 
significantly.

In bdbms, as a first step, we investigate the processing, e.g., indexing and querying, of Run-Length-Encoded (RLE) sequences. 
RLE~\cite{p10} is a compression technique that replaces the consecutive repeats of a character $C$ by one occurrence of $C$ followed by $C$'s frequency.  
One of the main challenges is how to operate on the compressed data without decompressing it. 
In~\cite{p5}, we proposed an index structure, termed the SBC-tree (\underline{\em S}tring \underline{\em B}-tree for \underline{\em C}ompressed sequences), 
for indexing and searching RLE-compressed sequences of arbitrary length. 
In Figure~\ref{fig:compression}, we illustrate how protein secondary structure sequences are stored in bdbms. We first compress the sequences using RLE, and 
then build an SBC-tree index over the compressed sequences. Queries over the sequences will use the index to retrieve the desired data without decompression.  
The SBC-tree is a two-level index structure based on the well-known 
String B-tree and a 3-sided range query structure. The SBC-tree supports substring as well 
as prefix matching, and range search operations over RLE-compressed sequences. The SBC-tree 
has an optimal external-memory space complexity as well as optimal search time for 
substring matching, prefix matching, and range search queries. 
More interestingly, SBC-tree has shown to be very practical to implement.
The SBC-tree index is prototyped in PostgreSQL with 
an R-tree in place of the 3-sided structure. Preliminary performance results illustrate 
that using the SBC-tree to index RLE-compressed protein sequences achieves up to an order 
of magnitude reduction in storage, up to 30\% reduction in I/Os for the insertion 
operations, and retains the optimal search performance achieved by the String B-tree over 
the uncompressed sequences.   

In bdbms, we plan to address the following challenges regarding the processing of compressed data:
\begin {itemize}
\item {\textbf{ Full integration  of the SBC-tree index:}
To fully integrate the SBC-tree index inside bdbms we plan to address several query processing and optimization issues including: 
(1) supporting subsequence matching, 
and (2) providing accurate cost functions for estimating the cost of the index. 
Subsequence matching is an important operation over biological sequences as it is used in many algorithms such as sequence alignment algorithms. 
We plan to extend the supported operations of the SBC-tree index to include subsequence matching.  
}

\item  {\textbf{Processing various formats of compressed data:} 
Currently, bdbms supports indexing and querying RLE-compressed sequence data. 
RLE is effective in the case of sequences where characters have long repeats in tandem. 
Compression techniques like 
gzip and Burrows-Wheeler Transform (BWT) can be more effective in compressing the other kinds of data. 
Our plan is to investigate indexing and querying other formats of compressed data in addition to RLE-compressed sequences
to efficiently support these data inside bdbms.
}
\end {itemize} 
\end{sloppypar}


\section{Related Work}
\label{sec:relatedwork}
\begin{sloppypar}
Periscope~\cite{p6, p7} is an ongoing project that aims at defining a declarative query language for querying biological data. 
Periscope/SQ~\cite{p7}, a component of Periscope, introduces new operators and data types that facilitate the processing and querying of sequence data. 
While the main focus of Periscope is on defining and supporting a new 
declarative query language, bdbms focuses on other functionalities that are required by biological databases, e.g., 
annotation and provenance management, local dependency tracking, update authorization, and non-traditional access methods.

Several annotation systems have been built to manage annotations over the web, e.g.,~\cite{p16, ref73, p18, ref74, ref75, p17}. 
Biodas (Biological Distributed Annotation System)~\cite{p16, p17} and Human Genome Browser~\cite{p18} are 
specialized biological annotation systems to annotate genome sequences. They allow users to integrate genome annotation information 
from multiple web servers. 
Managing annotations and provenance in relational databases has been addressed in~\cite{ref62, ref63, ref60, ref69, ref68, ref61}. 
In these techniques provenance data is pre-computed and stored inside the database as annotations. 
The main focus of these techniques is to propagate the annotations along with the query answer. 
Other aspects of annotation management, e.g., insertion, storage, and indexing, have not been addressed.   
Another approach for tracking provenance, termed the {\em lazy approach}, has been addressed in~\cite{ref66, ref65,  ref67, ref64}, 
where provenance data is computed at query time. 
{\em Lazy approach} techniques require that the derivation steps of the data to be known and to be invertible such that the provenance information can be computed. 
In bdbms, we treat provenance data as a kind of annotations because the derivation of biological data 
is usually ad-hoc and does not necessarily follow certain functions or queries. 

The access control and authorization process in current database systems is based on the GRANT/REVOKE model~\cite{ref71, ref70}. 
Although widely acceptable, this model lacks being content-based, i.e., the authorization is based only on the identity of the user. 
In bdbms, we propose the {\em content-based} approval model that is based on the data as well as on the identity of the user. 
\end{sloppypar}

\section{Concluding Remarks}
\label{sec:conclusion}
Two applications have been driving the bdbms project: 
building a database resource for the Escherichia coli (E.~coli) model organism and a protein structure database project.  
Through these two projects, we realized the need for the
functionalities that we address in bdbms, namely (1) Annotation and provenance management, 
(2) Local dependency tracking, (3) Update authorization, and (4) Non-traditional and novel access methods. 

bdbms is currently being 
prototyped using PostgreSQL.
In parallel work, we have extended relational algebra to operate on 
``annotated" relations.
The A-SQL language and the content-based authorization model are currently under development in PostgreSQL. 
The SP-GIST and SBC-tree access methods are already integrated inside PostgreSQL. 
We are currently studying several optimizations, cost estimates, 
and complex operations over these indexes.
  

\bibliographystyle{abbrv}
\bibliography{sigproc}  

\begin{thebibliography}{10}

\bibitem{p16}
biodas.org. http://biodas.org.

\bibitem{p22}
Exploiting the power of oracle using microsoft excel.
\newblock {\em Oracle White Paper}, December 2004.

\bibitem{p2}
W.~G. Aref and I.~F. Ilyas.
\newblock An extensible index for spatial databases.
\newblock In {\em Statistical and Scientific Database Management}, pages
  49--58, 2001.

\bibitem{p1}
W.~G. Aref and I.~F. Ilyas.
\newblock Sp-gist: An extensible database index for supporting space
  partitioning trees.
\newblock {\em Journal of Intelligent Information Systems}, 17(2-3):215--240,
  2001.

\bibitem{p8}
W.~Armstrong.
\newblock Dependency structures of database relationships.
\newblock In {\em International Federation for Information Processing (IFIP)},
  pages 580--583, 1974.

\bibitem{p13}
J.~L. Bentley.
\newblock Multidimensional binary search trees used for associative searching.
\newblock {\em Communications of the ACM}, 18(9):509--517, 1975.

\bibitem{ref62}
D.~Bhagwat, L.~Chiticariu, W.~Tan, and G.~Vijayvargiya.
\newblock An annotation management system for relational databases.
\newblock pages 900--911, 2004.

\bibitem{ref63}
P.~Buneman, A.~P. Chapman, and J.~Cheney.
\newblock Provenance management in curated databases.
\newblock In {\em {ACM SIGMOD International Conference on Management of Data
  }}, 2006.

\bibitem{ref66}
P.~Buneman, S.~Khanna, and W.-C. Tan.
\newblock Why and where: {A} characterization of data provenance.
\newblock {\em Lecture Notes in Computer Science}, 1973:316--333, 2001.

\bibitem{ref60}
P.~Buneman, S.~Khanna, and W.-C. Tan.
\newblock On propagation of deletions and annotations through views.
\newblock In {\em Principles of Database Systems (PODS)}, pages 150--158, 2002.

\bibitem{p11}
W.~A. Burkhard.
\newblock Hashing and trie algorithms for partial match retrieval.
\newblock {\em ACM Transactions Database Systems}, 1(2):175--187, 1976.

\bibitem{ref69}
L.~Chiticariu, W.-C. Tan, and G.~Vijayvargiya.
\newblock Dbnotes: a post-it system for relational databases based on
  provenance.
\newblock In {\em ACM SIGMOD International Conference on Management of Data},
  pages 942--944, 2005.

\bibitem{p9}
E.~Codd.
\newblock A relational model for large shared data banks.
\newblock In {\em Communications of the ACM 13:6}, pages 377--387, 1970.

\bibitem{ref65}
Y.~Cui and J.~Widom.
\newblock Practical lineage tracing in data warehouses.
\newblock In {\em {International Conference on Data Engineering}}, pages
  367--378, 2000.

\bibitem{ref67}
Y.~Cui and J.~Widom.
\newblock Lineage tracing for general data warehouse transformations.
\newblock In {\em {International Conference on Very Large Data Bases}}, pages
  471--480, 2001.

\bibitem{p4}
M.~Y. Eltabakh, R.~H. Eltarras, and W.~G. Aref.
\newblock Space-partitioning trees in postgresql: Realization and performance.
\newblock In {\em International Conference on Data Engineering}, pages
  100--111, 2006.

\bibitem{p5}
M.~Y. Eltabakh, W.-K. Hon, R.~Shah, W.~G. Aref, and J.~S. Vitter.
\newblock The sbc-tree: An index for run-length compressed sequences.
\newblock Technical Report CSD TR05-030, 2005.

\bibitem{ref71}
R.~Fagin.
\newblock On an authorization mechanism.
\newblock {\em ACM Transactions on Database Systems (TODS)}, 3(3):310--319,
  1978.

\bibitem{p15}
R.~A. Finkel and J.~L. Bentley.
\newblock Quad trees: A data structure for retrieval on composite keys.
\newblock {\em Acta Information}, 4:1--9, 1974.

\bibitem{p12}
E.~Fredkin.
\newblock Trie memory.
\newblock {\em Communications of the ACM}, 3(9):490--499, 1960.

\bibitem{ref68}
F.~Geerts, A.~Kementsietsidis, and D.~Milano.
\newblock Mondrian: Annotating and querying databases through colors and
  blocks.
\newblock In {\em International Conference on Data Engineering}, page~82, 2006.

\bibitem{p3}
T.~M. Ghanem, R.~Shah, M.~F. Mokbel, W.~G. Aref, and J.~S. Vitter.
\newblock Bulk operations for space-partitioning trees.
\newblock In {\em International Conference on Data Engineering}, pages 29--40,
  2004.

\bibitem{p10}
S.~W. Golomb.
\newblock Run-length encodings.
\newblock {\em IEEE Transactions on Information Theory}, 12:399--401, 1966.

\bibitem{ref70}
P.~P. Griffiths and B.~W. Wade.
\newblock An authorization mechanism for a relational database system.
\newblock {\em ACM Transactions on Database Systems (TODS)}, 1(3):242--255,
  1976.

\bibitem{p21}
H.~V. Jagadish and F.~Olken.
\newblock Database management for life sciences research.
\newblock {\em SIGMOD Record}, 33(2):15--20, 2004.

\bibitem{ref73}
J.~Kahan and R.~S. M.~Koivunen, E.~Prud'Hommeaux.
\newblock Annotea: An open rdf infrastructure for shared web annotations.
\newblock {\em WWW10}, pages 623--632, 2001.

\bibitem{p18}
W.~J. Kent, C.~W. Sugnet, T.~S. Furey, K.~M. Roskin, T.~H. Pringle, A.~M.
  Zahler, and D.~Haussler.
\newblock The human genome browser at ucsc.
\newblock {\em Genome Research}, 12(5):996--1006, 2002.

\bibitem{ref74}
D.~LaLiberte and A.~Braverman.
\newblock A protocol for scalable group and public annotations.
\newblock {\em WWW3}, pages 911--918, 1995.

\bibitem{p14}
R.~C. Nelson and H.~Samet.
\newblock A population analysis for hierarchical data structures.
\newblock In {\em ACM SIGMOD International Conference on Management of Data},
  pages 270--277, 1987.

\bibitem{p6}
J.~M. Patel.
\newblock The role of declarative querying in bioinformatics.
\newblock 7(1):89--92, 2003.

\bibitem{ref75}
M.~A. Schickler, M.~S. Mazer, and C.~Brooks.
\newblock Pan-browser support for annotations and other meta-information on
  theworld wide web.
\newblock {\em WWW5}, pages 1063--1074, 1996.

\bibitem{p17}
L.~Stein, S.~Eddy, and R.~Dowell.
\newblock Distributed sequence annotation system (das).
\newblock {\em Washigton University, Technical Report WUCS-01-07}, 2001.

\bibitem{ref58}
M.~Stonebraker, D.~Abadi, A.~Batkin, X.~Chen, M.~Cherniack, M.~Ferreira,
  E.~Lau, A.~Lin, S.~Madden, E.~O'Neil, P.~O'Neil, A.~Rasin, N.~Tran, and
  S.~Zdonik.
\newblock C-store: A column oriented dbms.
\newblock In {\em International Conference on Very Large Data Bases}, 2005.

\bibitem{p19}
M.~Stonebraker and G.~Kemnitz.
\newblock The postgres next generation database management system.
\newblock {\em Communications of the ACM}, 34(10):78--92, 1991.

\bibitem{ref61}
W.-C. Tan.
\newblock Containment of relational queries with annotation propagation.
\newblock In {\em International Symposium on Database Programming Languages},
  2003.

\bibitem{p7}
S.~Tata, J.~M. Patel, J.~S. Friedman, and A.~Swaroop.
\newblock Declarative querying for biological sequences.
\newblock In {\em International Conference on Data Engineering}, pages 87--96,
  2006.

\bibitem{p20}
T.~Topaloglou.
\newblock Biological data management: Research, practive and opportunities.
\newblock In {\em International Conference on Very Large Data Bases}, pages
  1233--1236, 2004.

\bibitem{ref64}
A.~Woodruff and M.~Stonebraker.
\newblock Supporting fine-grained data lineage in a database visualization
  environment.
\newblock In {\em {International Conference on Data Engineering}}, pages
  91--102, 1997.

\end{thebibliography}
\end{document}